\documentclass[10pt,prd,twocolumn,floatfix,tightenlines,showpacs,showkeys,preprintnumbers,nofootinbib, longbibliography]{revtex4-1}

\pdfoutput=1
\usepackage[colorlinks=true,breaklinks=true]{hyperref}
\hypersetup{allcolors=[rgb]{0.0 0.0 0.6},linkcolor=[rgb]{0.8 0.0 0.4}}
\usepackage{amsmath,amssymb,mathtools,tabu}
\usepackage{epsfig}  
\usepackage{graphicx}   
\usepackage{slashed}             
\usepackage{url}
\usepackage{color}
\usepackage{multirow}
\usepackage{placeins}
\usepackage[dvipsnames]{xcolor}
\usepackage{comment}
\usepackage{epsfig}
\usepackage{amsmath,bm,amsfonts}
\usepackage{hyperref}
\usepackage{multirow}
\usepackage{placeins}
\usepackage{color}
\usepackage{epsfig,verbatim}  
\usepackage{graphicx}                
\usepackage{url}
\usepackage{color}
\usepackage[dvipsnames]{xcolor}
\usepackage{slashed}
\allowdisplaybreaks

\bibpunct{[}{]}{,}{n}{}{,}

\newcommand{\w}{\omega}

\definecolor{purple}{rgb}{0.63,0.13,0.94}
\definecolor{red}{rgb}{1.0,0.0,0.0}
\definecolor{green}{rgb}{0.0,1.0,0.0}
\definecolor{blue}{rgb}{0.0,0.0,1.0}
\definecolor{orange}{rgb}{0.8,0.6,0,0}
\definecolor{magenta}{rgb}{0.8,0.0,0.6}
\definecolor{white}{rgb}{1.0,1.0,1.0}
\definecolor{black}{rgb}{0.0,0.0,0.0}

\begin{document}

\title{Fast Neutrino Flavor Conversion as Oscillations in a Quartic Potential}

\author{Basudeb Dasgupta}
\email{bdasgupta@theory.tifr.res.in}
\affiliation{Tata Institute of Fundamental Research, Homi Bhabha
Road, Mumbai 400005, India}

\author{Manibrata Sen}
\email{manibrata@theory.tifr.res.in}
\affiliation{Tata Institute of Fundamental Research, Homi Bhabha
Road, Mumbai 400005, India}

\date{December 15, 2017}
\preprint{TIFR/TH/17-34 }

\begin{abstract} 
Neutrinos in dense environments undergo collective pair conversions $\nu_e\bar{\nu}_e \leftrightarrow \nu_x\bar{\nu}_x$, where $x$ is a non-electron flavor, due to forward scattering off each other that may be a crucial ingredient for supernova explosions.  Depending on the flavor-dependent local angular distributions of the neutrino fluxes, the conversion rate can be ``fast,'' i.e., of the order $\mu=\sqrt{2}G_F n_\nu$, which can far exceed the usual neutrino oscillation frequency $\w=\Delta m^2/(2E)$. Until now, this surprising nonlinear phenomenon has only been understood in the linear regime and explored further using numerical experiments. We present an analytical treatment of the simplest system that exhibits fast conversions, and show that the conversion can be understood as the dynamics of a particle rolling down in a quartic potential, governed dominantly by $\mu$ but seeded by slower subleading effects.
\end{abstract}


\maketitle


\section{Introduction}
\label{sec:intro}

A core-collapse supernova (SN) offers perhaps the most extreme laboratory for studying neutrino flavor physics. While early studies focussed on vacuum oscillations and Mikheyev-Smirnov-Wolfenstein (MSW) matter effects~\cite{Wolfenstein:1977ue,Mikheev:1986gs}, deeper inside a supernova the neutrino density, $n_\nu$, is so large that non-linear neutrino-neutrino interactions can give rise to much more puzzling collective oscillations~\cite{Pantaleone:1992eq}.

These flavor oscillations, involving pair conversions of $\nu_e\bar{\nu}_e \leftrightarrow \nu_x\bar{\nu}_x$, where $\nu_x=\nu_\mu,\,\nu_\tau$, or any linear combination thereof, are collective in nature, i.e., all neutrino energies oscillate at the same frequency, and occur with a frequency $\sim \sqrt{\w\mu}$. Here, $\w=\Delta m^2/(2E)$ is the neutrino oscillation frequency in vacuum and $\mu=\sqrt{2}G_F n_\nu$ is the potential due to a neutrino density $n_\nu$. This collective frequency is much larger than $\w$ and could be the dominant mechanism of neutrino flavor conversion in supernovae. This has been a topic of heightened interest~\cite{Duan:2006an, Hannestad:2006nj, Fogli:2007bk,EstebanPretel:2007ec,Dasgupta:2009mg,Gava:2009pj,Banerjee:2011fj, Mirizzi:2011tu, Dasgupta:2011jf,Pejcha:2011en, Mirizzi:2012wp,Tamborra:2014aua,Chakraborty:2014lsa,Mangano:2014zda,Duan:2014gfa,Mirizzi:2015fva,Chakraborty:2015tfa,Dasgupta:2015iia,Abbar:2015fwa,Capozzi:2016oyk,Chakraborty:2016yeg,Das:2017iuj,Johns:2017oky,Dighe:2017sur} for the past decade, as reviewed in refs.~\cite{Duan:2010bg,Mirizzi:2015eza,Horiuchi:2017sku}.

There is still no analytical understanding of collective effects, in general, and much of our insight still comes from the simplest model that shows collective bipolar oscillations: a neutrino and an antineutrino beam interacting with each other. This system is mathematically equivalent to a pendulum in flavor space~\cite{Duan:2006an, Hannestad:2006nj}, similar to how the ordinary neutrino oscillations in vacuum or matter are equivalent to a precessing spin~\cite{Mikheev:1986gs,Stodolsky:1986dx,Kim:1987bv,Kim:1987ss}. Depending on the neutrino mass ordering, the gravitational force for this flavor pendulum acts upwards or downwards, thereby making certain flavor configurations unstable, akin to an inverted pendulum. Bipolar oscillations correspond to the pendulum starting in an unstable inverted position, slightly offset by a small mixing angle, and swinging through the lowest position to the other side. This mechanical analog of the flavor oscillations forms the basis for much of our intuitive understanding of the rich and puzzling physics of collective oscillations.
  
As early as 2005, it was claimed that even faster flavor conversions may occur in a SN~\cite{Sawyer:2005jk}. Such conversions, with a rate $\sim \mu \gg \sqrt{\w\mu} \gg \w$, seem to require nontrivial flavor-dependent angular distributions for neutrinos and antineutrinos. This was further studied in refs.~\cite{Sawyer:2008zs, Sawyer:2015dsa, Chakraborty:2016lct, Dasgupta:2016dbv,Sen:2017ogt,Izaguirre:2016gsx, Capozzi:2017gqd} and, as an end product of these studies, it was concluded that one requires a crossing in the electron lepton number intensities to obtain a gap in the dispersion relation for modes of flavor evolution, which leads to convective or absolute instabilities that causes fast flavor conversion. This condition is quite similar to how spectral crossings are needed for the development of the bipolar instability modes~\cite{Dasgupta:2009mg}. 

A major conceptual gap in the understanding of fast conversions is that fast oscillations have never been studied analytically in the fully nonlinear regime. As a result, one doesn't understand \emph{why} do the fast oscillations take place. This is the gap that we will fill in this paper. 

Our aim is to discover the mechanical analog of fast oscillations, roughly analogous to how the flavor pendulum explains bipolar flavor oscillations. Towards this goal, we consider the simplest model that shows fast oscillations and, under some simplifying assumptions, show that its dynamics is equivalent to that of a particle in a quartic potential. Fast oscillations correspond to the inversion of this potential, leading to an instability. Using the classical mechanical action, we analytically compute the oscillation period in the inverted quartic potential and find agreement with numerical solutions, both for constant and varying neutrino-induced potential $\mu$. We further explore this problem, analytically as far as tractable, to identify the exactly and approximately conserved quantities, and to provide semi-quantitative understanding for two out of the three different time-scales in the problem. We begin our analysis below.

\section{Fast oscillations in 4 beam model}
\label{sec:2}
The equation of motion (EoM) for a 2-flavor neutrino of momentum ${\bf p}$, represented by a 3-component Bloch vector is given by,
\begin{equation}\label{EoM}
\dot{\bf P}_{\bf p}=\bigl[\omega_{{\bf p}}{\bf B} +\mu\int d\Gamma'(1-{\bf v}\cdot{\bf v'}){\bf P}_{\bf p'}\bigr]\times{\bf P}_{\bf p}\,,
\end{equation}
where ${\bf B}=(\sin2\vartheta_0,\,0,\,-\cos2\vartheta_0)$ for a vacuum mixing angle $\vartheta_0$, and $d\Gamma'$ refers to an integral over the 3-momenta of the other neutrinos. Here, we have ignored ordinary matter effects and assumed that the above gas of collisionless neutrinos is homogeneous over a length scale much larger than the length scale corresponding to fast conversions, and thus the only relevant dynamics is its time evolution. Similar equations hold for antineutrinos with the replacement $\overline{{\bf P}}_{\w_{\bf p},{\bf v}_{\bf p}}\equiv{\bf P}_{-\w_{\bf p},{\bf v}_{\bf p}}$. In the following, we drop the subscript ${\bf p}$ for clarity. 

\begin{figure}[!t]
\includegraphics[width=0.85\columnwidth]{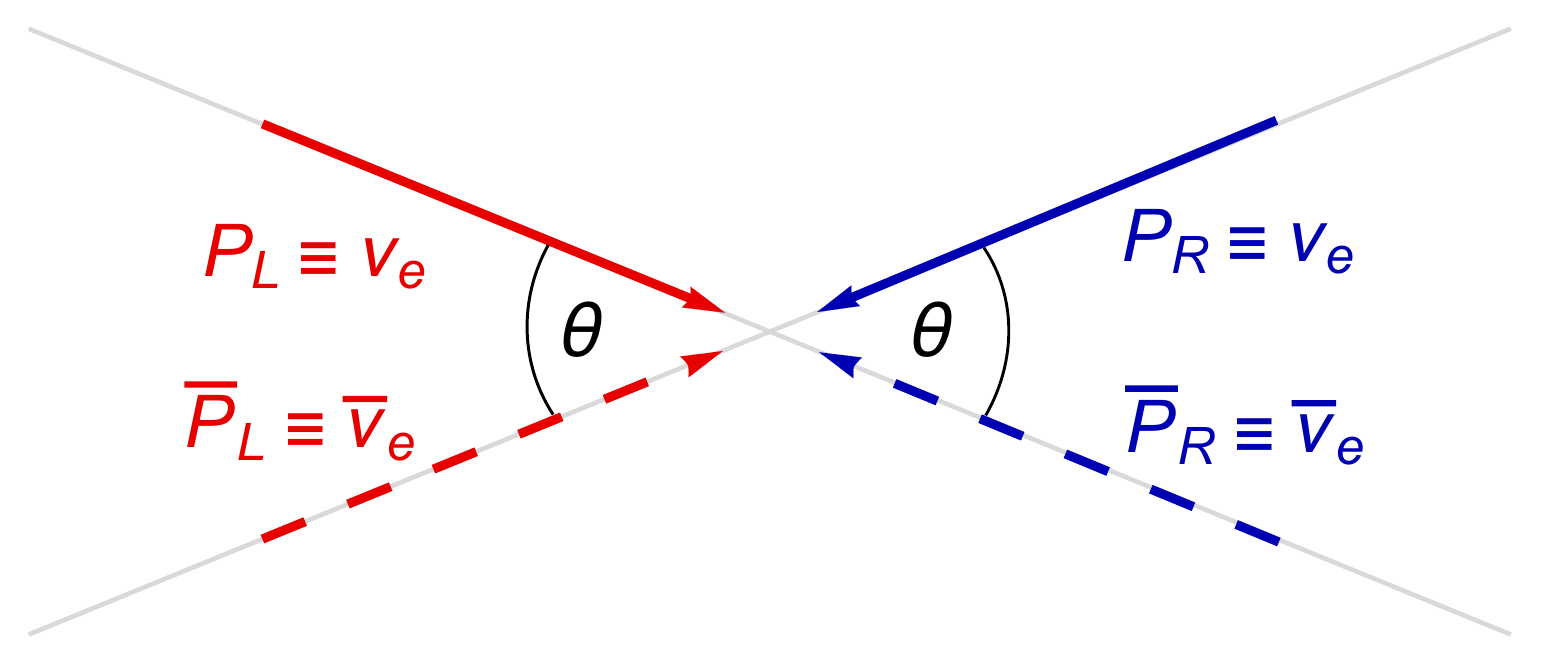}
\caption{ Four-beam model: Electron neutrinos (solid) and antineutrinos (dashed) travelling along two beams each, one from the left (red) and another from the right (blue), forward scatter off each other. We study the time evolution of the flavor content of these beams.}
\label{fig1}
\end{figure} 

The simplest system that shows fast flavor conversions is a set of four beams of neutrinos and antineutrinos intersecting each other as shown in Fig.\,\ref{fig1} and governed by Eq.(\ref{EoM}). The terms involving ${\bf v}\cdot{\bf v'}$ lead to terms involving $c\equiv\cos\theta$, where $\theta$ is the angle shown in Fig.\,\ref{fig1}. The flavor evolution is more clearly understood in terms of the following linear combinations of the polarization vectors,
\begin{eqnarray}
\label{redef1}{\bf Q}   &\equiv& {\bf P}_{ L}+{\bf P}_{ R}+\overline{{\bf P}}_{ L}+\overline{{\bf P}}_{ R}-\tfrac{2\w}{\mu(3-c)}{\bf B}\,,\\
\label{redef2} {\bf D}   &\equiv& {\bf P}_{ L}+{\bf P}_{ R}-\overline{{\bf P}}_{ L}-\overline{{\bf P}}_{ R}\,,\\
\label{redef3} {\bf X} &\equiv& {\bf P}_{ L}-{\bf P}_{ R}+\overline{{\bf P}}_{ L}-\overline{{\bf P}}_{ R}\,,\\
\label{redef4} {\bf Y} &\equiv& {\bf P}_{ L}-{\bf P}_{ R}-\overline{{\bf P}}_{ L}+\overline{{\bf P}}_{ R}\,,
\end{eqnarray}
in terms of which the EoMs take the form
\begin{eqnarray} 
 \label{4-modeQD1}\dot{\bf Q}     &=& \frac{\mu}{2}(3-c)\,{\bf D}\times{\bf Q}+\frac{\mu}{2}(1+c)\,{\bf X}\times{\bf Y}\,,\\
 \label{4-modeQD2}\dot{\bf D}     &=& \w\, {\bf B}\times{\bf Q}\,,\\
 \label{4-modeQD3}\dot{\bf X}      &=& \biggl[\w\left(\frac{3+c}{3-c}\right) \,{\bf B}+\mu\,c\,{\bf Q}\biggr]\times{\bf Y}+\mu\,{\bf D}\times{\bf X}\,,\\
\label{4-modeQD4} \dot{\bf Y}      &=&\biggl[\w\left(\frac{2}{3-c}\right) \,{\bf B}-\frac{\mu}{2}\,(1-c){\bf Q}\biggr]\times{\bf X} \nonumber\\
                         &&+\frac{\mu}{2}(3+c)\,{\bf D}\times{\bf Y}\,.
\end{eqnarray}

\subsection{Bipolar limit}
There are two ways in which the above set of equations reduce to the previously well-known equations for the bipolar flavor pendulum, e.g., in refs.~\cite{Duan:2006an, Hannestad:2006nj}. Firstly, if $c=-1$ then Eqs.(\ref{4-modeQD1},\,\ref{4-modeQD2}) decouple from the rest and simply reproduce the bipolar flavor pendulum. In this limit, Eqs.(\ref{4-modeQD3},\,\ref{4-modeQD4}) imply that ${\bf X}\cdot{\bf X}+{\bf Y}\cdot{\bf Y}$ is constant, and if ${\bf X}$ and ${\bf Y}$ are initially zero, they remain zero. Secondly, for any value of $c$, if ${\bf X}$ and ${\bf Y}$ are initially exactly zero, i.e., there is a $L\leftrightarrow R$ exchange symmetry in Eqs.(\ref{redef3},\,\ref{redef4}), they do not evolve at all. This is to be expected because the equations of motion do not break this symmetry unless the initial conditions do so. In this case, the first two equations simply reproduce the flavor pendulum that exhibits bipolar oscillations at a frequency $\sim\sqrt{\w\mu}$. In addition, if the initial neutrino-antineutrino asymmetry $\alpha$, defined such that $\overline{P}_z=(1-\alpha) P_z$ is zero, the ${\bf Q}$ only evolves in the $x$-$z$ plane while ${\bf D}$ acquires a non-zero component only along the $y$ direction. Here we take $0\leq \alpha\leq 1$ and $|{\bf P}|=1$, corresponding to an excess of neutrinos over antineutrinos as is expected in SNe. On the other hand, if there is an excess of antineutrinos, it is more convenient to define  $P_z=(1-\bar{\alpha})\overline{P}_z$ with $0\leq \bar{\alpha}\leq 1$ and $|\overline{{\bf P}}|=1$. If $\alpha\,\,{\rm or\,\,}\bar{\alpha}\neq0$, the pendulum has a spin that makes it gyrate like a top~\cite{Hannestad:2006nj}.

\subsection{Fast oscillations beyond the bipolar limit}
It is thus clear, as was already evident through the linear analysis in ref.~\cite{Chakraborty:2016lct}, that one must break the $L\leftrightarrow R$ symmetry to obtain any oscillations faster than the bipolar oscillations. We will consider initial conditions on the polarization vectors to be\begin{eqnarray}\label{inCond}
  {\bf P}_{L,R}(0) &=& \left(0,\,0,\,1\pm\epsilon\right)\,,\\
  \overline{{\bf P}}_{L,R}(0)&=&\left(0,\,0,\,1-\alpha\pm\epsilon\right)\,,
\end{eqnarray}
where $\alpha$ parametrizes the asymmetry between neutrino and antineutrino number densities and $\epsilon$ is the small difference between the left and right going modes that breaks the $L\leftrightarrow R$ symmetry. In general the motion is quite complicated but for the above initial conditions and $\alpha=0$, ${\bf Y}$ is in the $y$ direction only and ${\bf X}$ remains in the $x$-$z$ plane. One can see this by inspecting Eqs.(\ref{4-modeQD1}\,-\,\ref{4-modeQD4}). This $\alpha=0$ limit is significantly simpler and we confine our attention to it to illustrate the physics of fast oscillations. Many of the obtained insights will be relevant more generally.

\begin{figure}[!t]
\includegraphics[width=0.75\columnwidth]{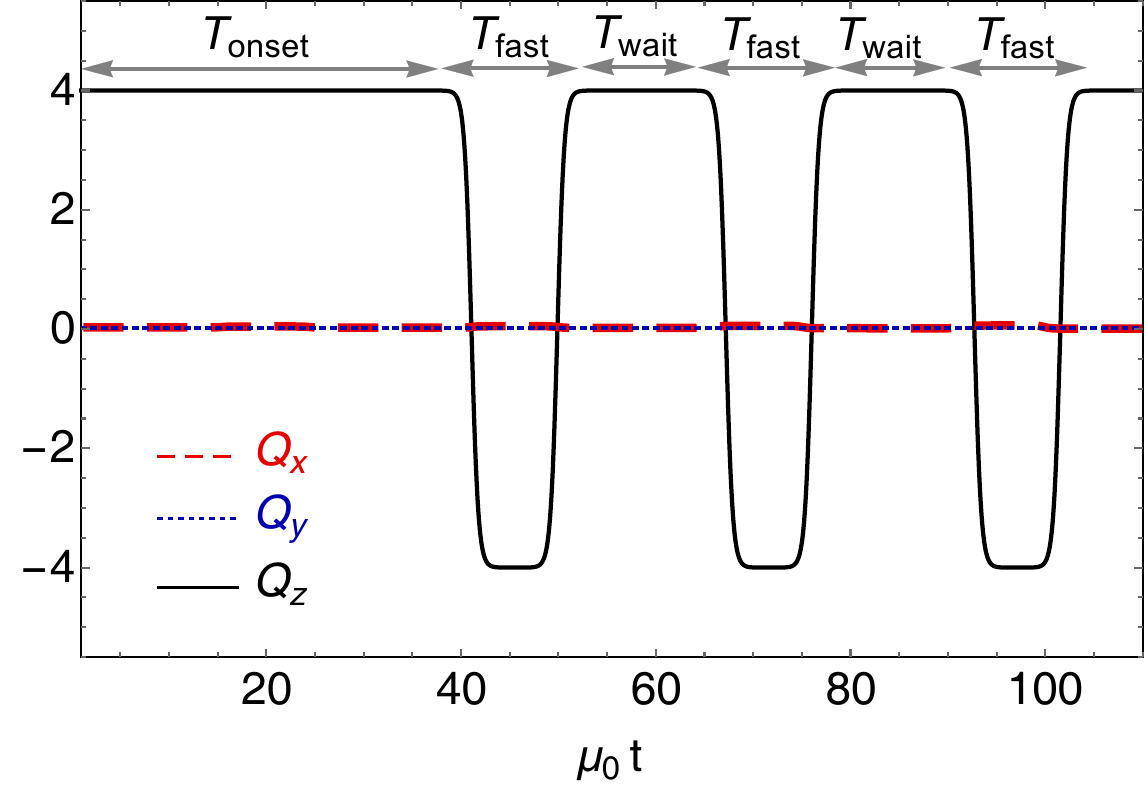}
\caption{Dynamics of the components of ${\bf Q}$. The parameters are chosen to be $\w/\mu_0=10^{-5}$, $\vartheta_0=10^{-2}$ and $c=0.5$. Here $\mu=\mu_0=10^{5}\,{\rm km}^{-1}$ is the value of $\mu$ at the neutrinosphere. $T_{\rm fast}$ is matched using the estimate in Eq.(\ref{Tfast}), which defines $T_{\rm onset}$ and $T_{\rm wait}$ as the periods where $Q_z\geq0.99\,Q_z(0)$.}
\label{fig2}
\end{figure}

\subsubsection{Conserved quantities}

We now identify the conserved quantities. Eq.(\ref{EoM}) implies that the magnitudes of each of the 4 polarization vectors ${\bf P}_{\bf p}$ remains constant. Further, Eq.(\ref{4-modeQD2}) provides that ${\bf B}\cdot{\bf D}$ is a constant of motion, as in the bipolar case. This proves conservation of flavor lepton number even for fast oscillations, as one would expect. 

The length of ${\bf Q}$, unlike for bipolar oscillations, is not conserved and changes as
\begin{equation}
  \frac{d}{dt}\left({\bf Q}\cdot{\bf Q}\right)=\mu\frac{(1+c)}{2}\bigl[{\bf Q}{\bf X}{\bf Y}\bigr] \,,
\end{equation}
where $[\cdots]$ indicates the scalar triple product of the three vectors. The evolution of the components of ${\bf Q}$ is shown in Fig.\,\ref{fig2}. The dynamics is mainly captured in $Q_z$, with $Q_x,\,Q_y\simeq0$.

\begin{figure}[!t]
\includegraphics[width=0.40\textwidth,height=0.26\textwidth]{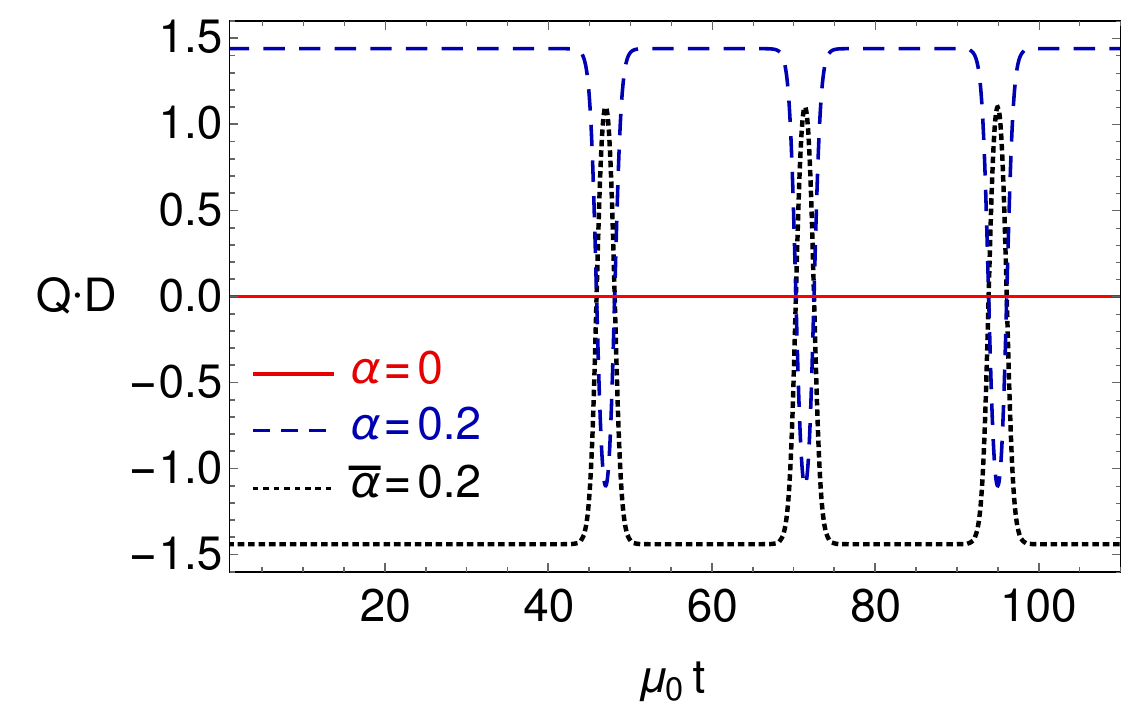}
\caption{Variation of ${\bf Q}\cdot{\bf D}$ with time for neutrino-antineutrino asymmetry $\alpha=0$ (solid red), $\alpha=0.2$ (dashed blue) and $\bar\alpha=0.2$ (dotted black).}
\label{fig3}
\end{figure}

Likewise, the quantity ${\bf Q}\cdot{\bf D}$ varies as 
 \begin{equation}\label{QD}
  \frac{d}{dt}\left({\bf Q}\cdot{\bf D}\right)=\mu\frac{(1+c)}{2}\bigl[{\bf D}{\bf X}{\bf Y}\bigr] \,.
 \end{equation}
 If there is no initial asymmetry, i.e., $\alpha=0$ and therefore ${\bf D}(0)=0$, the r.h.s. of Eq.(\ref{QD}) vanishes because ${\bf D}$ and ${\bf X}\times{\bf Y}$ remain orthogonal, as we argued following Eqs.\mbox{(\ref{4-modeQD1}\,-\,\ref{4-modeQD4})}. Then, ${\bf Q}\cdot{\bf D}$ is a constant and remains at its initial value zero. However, for $\alpha \neq 0$, i.e., a non-zero neutrino-antineutrino asymmetry, ${\bf Q}\cdot{\bf D}$ is no longer constant, unlike for the bipolar flavor pendulum~\cite{Hannestad:2006nj}.While a core-collapse SN mostly has an excess of neutrinos over antineutrinos, in the recently discovered lepton-emission self-sustained asymmetry (LESA) phenomenon \cite{Tamborra:2014aua} as well as in binary neutron star mergers \cite{Frensel:2016fge,Wu:2017qpc,Tian:2017xbr}, there can be an excess of antineutrinos over neutrinos, leading to a non-zero value of $\bar\alpha$. In Fig.\,\ref{fig3}, we show ${\bf Q}\cdot{\bf D}$ for $\alpha=0$ as well as for $\alpha=0.2$ and $\bar\alpha=0.2$. Defining $\bar{\alpha}$, instead of simply letting $\alpha$ be negative, has the advantage that $\alpha=0.2$ and $\bar\alpha=0.2$ are related to each other very simply as is apparent from Fig.\,\ref{fig3}. In the limit $\w\rightarrow0$, the replacement ${\bf P}\leftrightarrow\overline{{\bf P}}$ keeps the EoMs unchanged.

As an immediate by-product, one can solve for ${\bf D}$ starting from Eq.(\ref{4-modeQD1}). Taking a cross product with ${\bf Q}$, one gets
\begin{eqnarray}
 {\bf D} &=& \frac{2}{\mu(3-c)}\frac{{\bf Q}\times\dot{\bf Q}}{{\bf Q}^2}+\frac{{\bf Q}\cdot{\bf D}}{{\bf Q}^2}{\bf Q} \,\nonumber\\
  && + \frac{(1+c)}{(3-c)}\frac{1}{{\bf Q}^2}\bigl[({\bf Q}\cdot{\bf X}){\bf Y}-({\bf Q}\cdot{\bf Y}){\bf X}\bigr] \,.
\end{eqnarray}
The terms on the first line are identical to the ${\bf D}$ for the bipolar pendulum~\cite{Hannestad:2006nj}, but one must note that ${\bf Q}$ obeys a different equation than in the bipolar oscillations. Thus, even if the terms on the second line are small (they indeed are), the solution for ${\bf D}$ is actually different! Moreover, ${\bf Q}\cdot{\bf D}$ is not constant if $\alpha\neq 0$, and this expression for ${\bf D}$ must be understood as an implicit solution.

\subsubsection{Conserved quantities in the limit $\w/\mu\to0$ and $\alpha=0$}

In addition to the above conditionally but exactly conserved quantities, there are some approximately conserved quantities. In the limit of large neutrino interactions, i.e., $\w/\mu\to0$, Eq.(\ref{4-modeQD2}) gives that ${\bf D}$ is a constant. If further $\alpha=0$, then ${\bf D}$ can be set to zero. This simplifies Eqs.(\ref{4-modeQD1}\,-\,\ref{4-modeQD4}) immensely, giving
\begin{eqnarray} 
\label{4-modeD01} \dot{\bf Q}                    &=& \frac{\mu}{2}(1+c)\,{\bf X}\times{\bf Y}\,,\\
\label{4-modeD02} \dot{\bf X}            &=& \mu\,c\,{\bf Q}\times{\bf Y} \,,\\
\label{4-modeD03}\dot{\bf Y}                 &=&-\frac{\mu}{2}(1-c)\,{\bf Q}\times{\bf X}\,.
\end{eqnarray}
One then immediately finds that ${\bf Q}\cdot{\bf X},\,{\bf Q}\cdot{\bf Y},$ and ${\bf X}\cdot{\bf Y}$, as well as $2c\,{\bf Q}\cdot{\bf Q}+{(1+c)}\,{\bf X}\cdot{\bf X}$ and $(1-c)\,{\bf Q}\cdot{\bf Q}+{(1+c)}\,{\bf Y}\cdot{\bf Y}$ are conserved in this limit. 

Differentiating Eq.(\ref{4-modeD01}), one finds
\begin{equation}
\label{ModeQ1}
 \ddot{\bf Q}        = -\mu^2\,c\,(1-c)\biggl[|{\bf Q}_0|^2-{\bf Q}\cdot{\bf Q}\biggr]{\bf Q}\,, 
\end{equation} 
which is a closed equation for ${\bf Q}$ that derives from the Lagrangian 
\begin{equation}
\label{ModeQ2}
 \mathcal{L}_{\bf Q} = \frac{1}{2}|\dot{\bf Q}|^2-\,\mu^2\,c\,(1-c)\biggl[|{\bf Q}_0|^2-\frac{{\bf Q}\cdot{\bf Q}}{2}\biggr]\frac{{\bf Q}\cdot{\bf Q}}{2}\,,                   
\end{equation}
where $|{\bf Q}_0|$ is the modulus of ${\bf Q}$ at time $t=0$. Using Eq.(\ref{ModeQ1}) one finds the total energy is
\begin{equation}
\label{EEE}
  E= \frac{1}{2}|\dot{\bf Q}|^2+\,\mu^2\,c\,(1-c)\biggl[|{\bf Q}_0|^2-\frac{{\bf Q}\cdot{\bf Q}}{2}\biggr]\frac{{\bf Q}\cdot{\bf Q}}{2}\,,
\end{equation}
which is an additional constant of motion. Note that ${\bf Q}$ is confined to the $x$-$z$ plane when $\alpha=0$, and $Q_x$ can be eliminated using $E$, thereby reducing the problem to the study of only the $z$ component of ${\bf Q}$ to understand the flavor evolution shown in Fig.\,\ref{fig2}. Clearly, as $Q_x\simeq {\cal O}(\vartheta_0)$, the energy $E$ is dominated by $Q_z$.

\begin{figure}[!t]
\includegraphics[width=0.44\textwidth,height=0.25\textwidth]{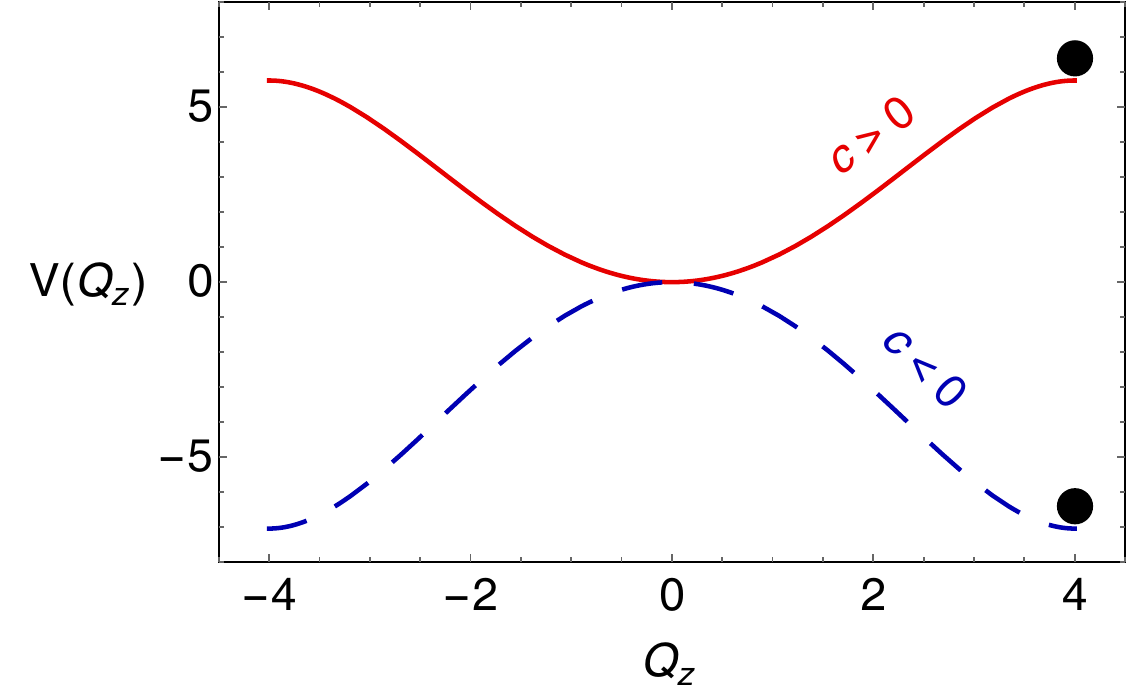}
\caption{Left: The potential $V(Q_z)$ for two different values of  $c=0.1$ (solid red) and  $c=-0.1$ (dashed blue).}
\label{fig4}
\end{figure}

\subsubsection{Particle in a quartic potential}
An interesting feature is that fast conversions exist only for certain angular distributions of the neutrino
beams. Using a linear stability analysis, it was shown in ref.~\cite{Chakraborty:2016lct} that fast conversions exist only for $c\equiv\cos\theta>0$. The reason for this becomes obvious if one observes the potential term $V(Q_z)$ in $\mathcal{L}_{\bf Q}$. Classically, this relates to motion of a particle in a quartic potential given by 
\begin{equation}
 V( Q_z)\approx\,\mu^2\,c\,(1-c)\biggl[|{\bf Q}_0|^2-\frac{Q_z^2}{2}\biggr]\frac{Q_z^2}{2}\,.
\label{potential}
\end{equation}
As shown in Fig.\,\ref{fig4}, the potential is an inverted quartic for $c<0$ and a quartic for $c>0$. The motion of $ Q_z$ is governed by this potential. Given the initial condition $Q_z(0)=4\left[1-(\w\,\cos2\vartheta_0)/(2\mu\,(3-c))\right]$, for $c>0$ the potential causes $Q_z$ to roll down towards the bottom of the potential well and subsequently oscillate in it. 
In flavor space, these are fast conversions. On the other hand, for $c<0$ a potential barrier is encountered by $Q_z$. The value of $Q_z$ therefore remains at its initial value and there are no fast conversions. 
Note that the above initial condition for $Q_z$ is for the inverted mass ordering, where $\w<0$. For normal mass ordering, the same initial condition holds with the replacement $\w\rightarrow -\w$. However, fast conversions are essentially independent of the mass ordering. In fact, even the triggering of fast conversions, that is dependent on $\w$, does not seem to crucially depend on the sign of $\w$.

\begin{figure}[!t]
\includegraphics[width=0.4\textwidth,height=0.25\textwidth]{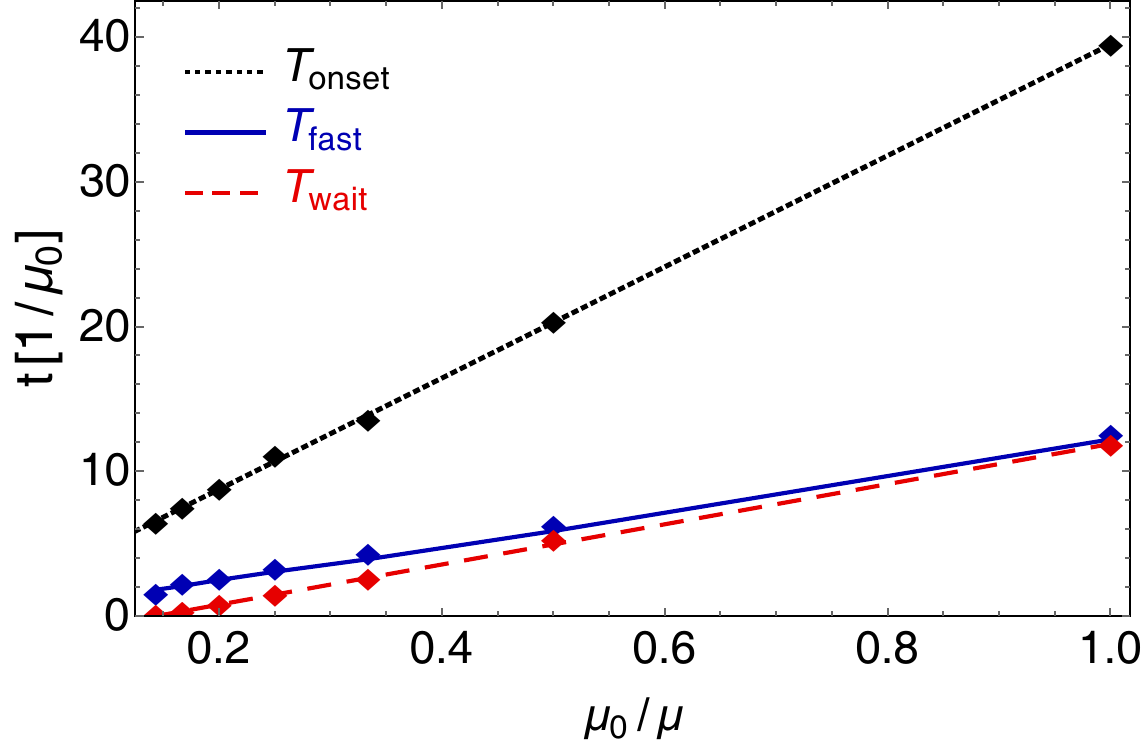}
\caption{Time periods $T_{\rm onset},\,T_{\rm fast}$ and $T_{\rm wait}$ and their linear dependence on $1/\mu$. Dots show the the numerical data whereas the lines are the best fit through them. While the fit for $T_{\rm fast}$ is given from Eq.(\ref{Tfast}), those for  $T_{\rm onset}$ and $T_{\rm wait}$ are obtained numerically.}
\label{fig5}
\end{figure}

In order to verify whether the above analytical approximations explain the evolution of ${\bf Q}$, we numerically solved Eqs.(\ref{4-modeQD1}\,-\,\ref{4-modeQD4}) and compared with the numerical solution of Eq.(\ref{ModeQ1}). These results are shown in Fig.\,\ref{fig2}. One observes that there are three timescales: $T_{\rm onset}$, the onset time; $T_{\rm fast}$, characterizing the time-period of fast oscillations; and $T_{\rm wait}$, the waiting period in between two oscillations.
We do not expect Eq.(\ref{ModeQ1}) to give the correct solution at initial times up to $T_{\rm onset}$ and in between the oscillations for the periods designated $T_{\rm wait}$. For these periods, roughly ${Q_z}\gtrsim 0.99\,Q_z(0)$ and the r.h.s. of Eq.(\ref{ModeQ1}) is very small, i.e., $\lesssim{\cal O}(\omega/\mu)=10^{-5}$. Thus the flavor evolution is governed by the $\w$-dependent and otherwise sub-dominant terms which we have ignored (see Appendix\,\ref{sec:App}). On the other hand, in this regime, the solution is already very well understood using linear stability analysis. More interestingly, the evolution of $Q_z$ is very well explained using Eq.(\ref{ModeQ1}) when it is strongly nonlinear, i.e., deviates appreciably from its initial value.

One can compute the time-period of the fast oscillations using energy conservation, to get
\begin{equation}
 T_{\rm fast}=2\int_{{Q_{z}^{\rm max}}}^{{Q_{z}^{\rm min}}}\,\frac{d Q_z}{\sqrt{2\big(E-V(Q_z)\big)}}\,.
\label{Tfast}
\end{equation}
This integral is in fact analytically expressible in terms of an elliptic function. However, the result is opaque and lengthy and we don't display it here. Evaluating the same, we find that it matches quite well with the numerical results shown in Fig.\,\ref{fig5}, if we consider $Q_z^{\rm max}\approx 0.99\, Q_z(0)$. The blue dots represent the fast time-period (excluding the onset and waiting times, as previously noted) obtained from numerical solution of Eqs.\mbox{(\ref{4-modeQD1}\,-\,\ref{4-modeQD4})}, whereas the solid blue line is obtained by evaluating the integral in Eq.(\ref{Tfast}). 

\begin{figure}[!t]
\includegraphics[width=0.81\columnwidth]{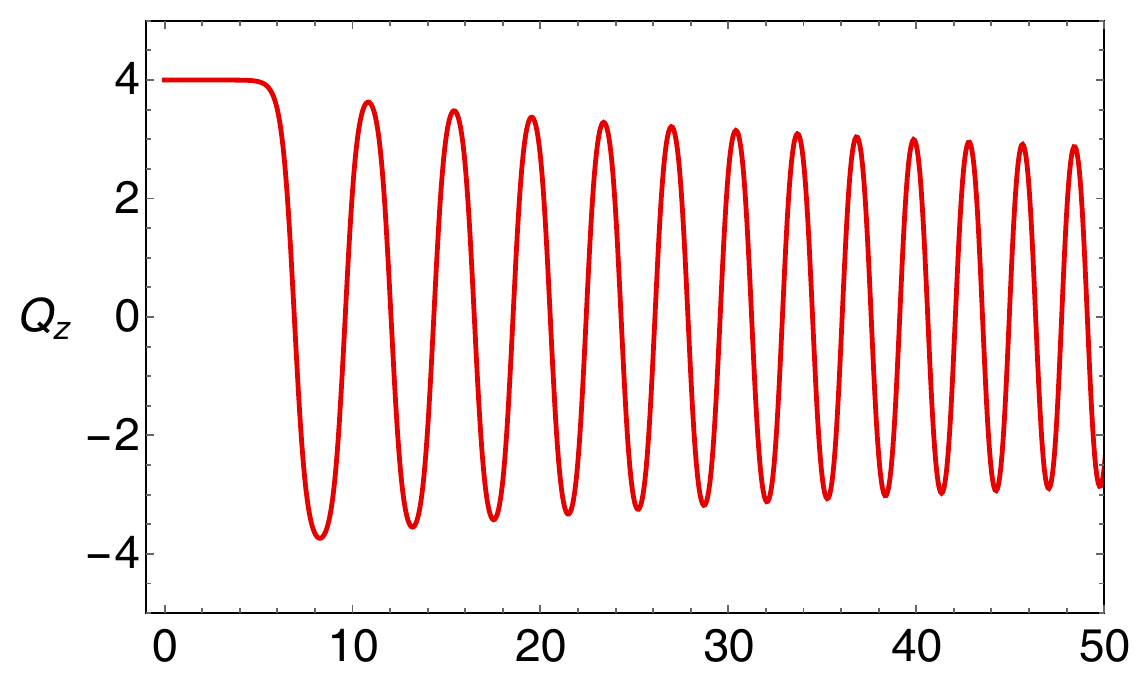}\vspace{0.2cm}\\
\includegraphics[width=0.81\columnwidth]{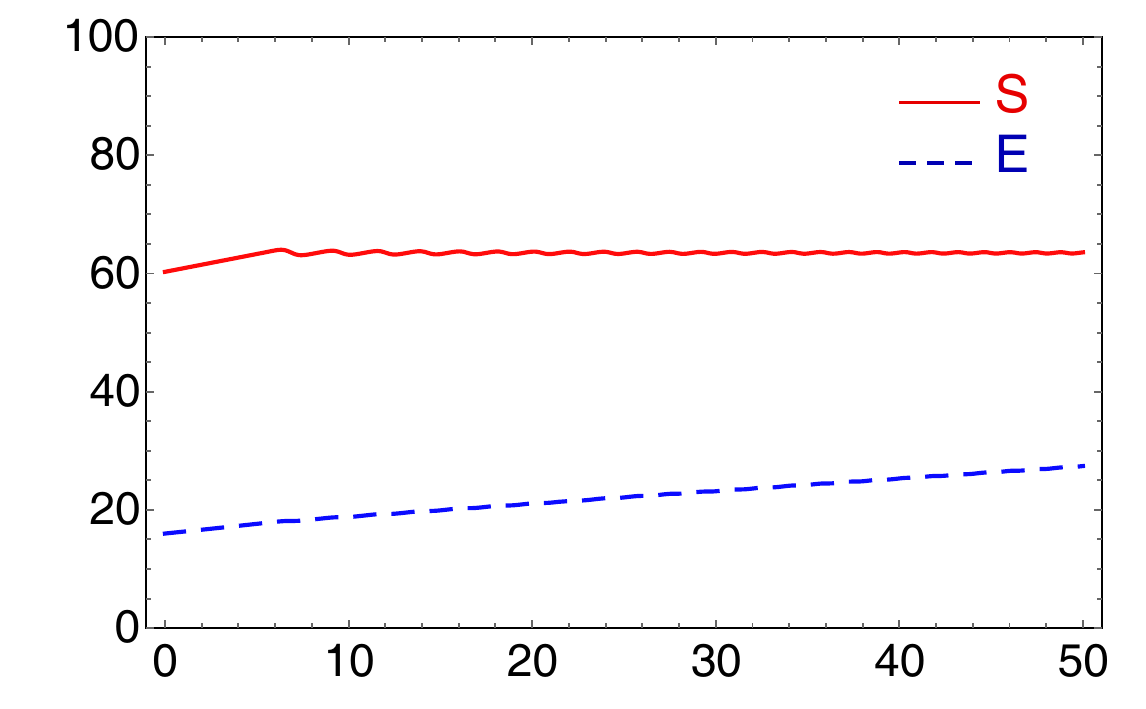}\vspace{0.2cm}\\
\includegraphics[width=0.81\columnwidth]{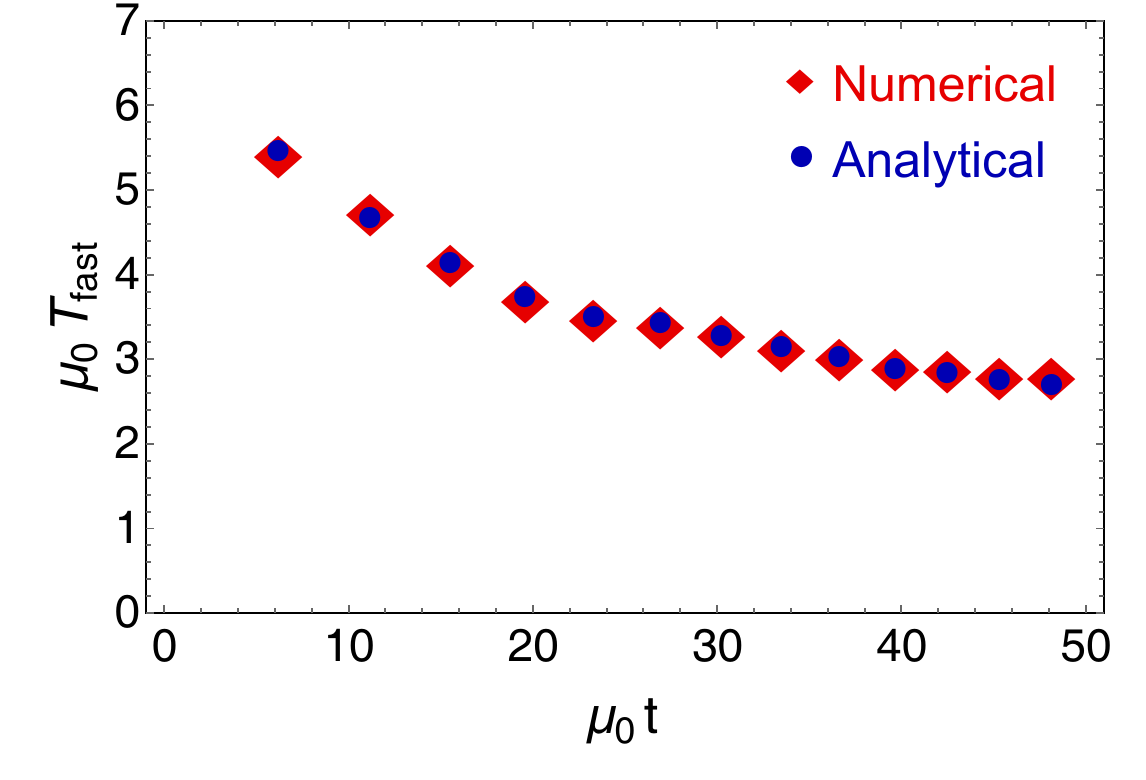}
\caption{Top: Variation of $Q_z$ for a time-varying neutrino-neutrino potential given by $\mu(t)=\mu_0(1+t/100)$. Middle: Plot of the action $S$ and the energy $E(t)$. Note how the energy changes, but action remains constant. Bottom: Variation of $T_{\rm fast}$ with time.}
\label{fig6}
\end{figure}

\begin{figure}[!b]
\hspace{-0.2cm}\includegraphics[width=0.37\textwidth]{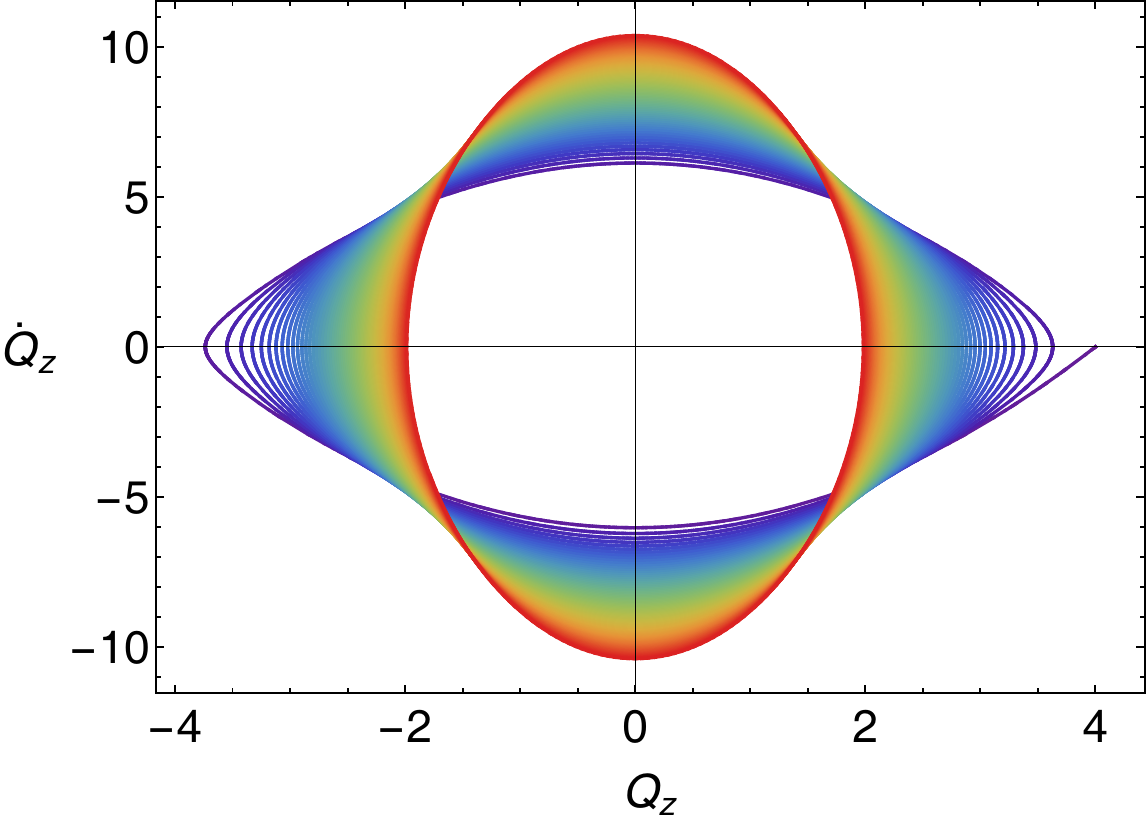}
\caption{Trajectory in phase space for varying $\mu$. Redder colors refer to later times and larger $\mu$.}
\label{fig7}
\end{figure}

Now we briefly discuss what happens if $\mu$ is not a constant, but rather varies with time as $\mu(t)$. One expects that if $\mu(t)$ is time-dependent, the energy $E(t)$ also becomes time-dependent. Naturally, the time period $T_{\rm fast}$ also changes with time. In Fig.\,\ref{fig6}, we show the evolution of $Q_z$ (top panel) for a time-dependent neutrino potential $\mu(t)=\mu_0(1+t/100)$.

While this is in general a much more complicated problem, if the rate of change of $\mu(t)$ is much smaller than the frequency of fast oscillations (as chosen above) one can use adiabatic invariance to derive some simple results. In the adiabatic limit, the action variable of the system
\begin{equation}
\label{action}
 S(E,\mu)=\oint p_Q\,dQ_z=\oint \sqrt{2\bigl(E-V(Q_z)\bigr)}\,dQ_z\,,
\end{equation}
remains invariant to a good approximation. Here the generalized momentum for the system is $p_Q=\dot{Q}_z$, neglecting $Q_x\simeq0$. This action $S(E(t),\mu(t))$ remains invariant under adiabatic changes in $\mu(t)$ while energy changes appreciably, as shown in the middle panel of Fig.\,\ref{fig6}.  In Fig.\,\ref{fig7}, we show the phase trajectory for the time-varying $\mu(t)$ above. As $\mu(t)$ increases with time, the potential becomes deeper and the oscillation amplitude decreases but the energy increases; the closed trajectory in phase space becomes more oblong along momentum, keeping the enclosed area constant.

It is possible to analytically perform the integral in Eq.(\ref{action}), giving a closed expression for the adiabatic invariant $S$ in terms of $E(t)$ and $\mu(t)$. One can then compute an analytical expression for the time-dependent time-period $T_{\rm fast}(t)$, using
\begin{equation}
 T_{\rm fast}(t)=\frac{\partial}{\partial E}\,S\bigl(E,\mu(t)\bigr)\,.
\end{equation}
As the expressions are unwieldy, and easily reproduced, we omit them here.
In the bottom panel of Fig.\,\ref{fig6}, we show the time period computed analytically in this manner (blue dots), compared with the same measured from the numerical solutions of the EoMs (red dots). This is based on a single calibration between our analytical estimate of $T_{\rm fast}$ and the numerics that we used to identify $Q_z^{\rm max}=0.99\,Q_z(0)$ as the boundary where the slower terms become dominant. Subsequently, this agreement at different and changing $\mu$ highlights that the agreement is not superfluous or accidental.

The other two time scales, $T_{\rm onset}$ and $T_{\rm wait}$, are somewhat harder to estimate. We have checked numerically that all of them vary as $1/\mu$, as seen in Fig.\,\ref{fig5}. In addition, we find that $T_{\rm onset}$ depends logarithmically on the ``seed'' given in Eq.(\ref{4-modeQD1}). Solving Eq.(\ref{ModeQ1}) for $Q_z$, and determining $T_{\rm onset}$ by checking for small deviations of $Q_z$ from its initial value gives,
\begin{equation}
 T_{\rm onset}\propto\frac{1}{\mu\sqrt{2c(1-c)}}\,{\rm ln}\biggl[\frac{(3-c)}{\cos2\vartheta_0}\,\frac{\mu_0}{\w}\biggr]\,,
\end{equation}
which underestimates $T_{\rm onset}$ by approximately a factor of 2, relative to the numerical value seen in Fig.\,\ref{fig2}. For $T_{\rm wait}$ as well, we find numerically that it depends logarithmically on $\vartheta_0$ and $\w$. More detailed numerical evidence for these logarithmic dependences is presented in Appendix\,\ref{sec:App}.

\subsubsection{Asymmetric fast oscillations}
We now turn to the case when the initial neutrino-antineutrino asymmetry is nonzero, i.e., $\alpha\neq0$. Examining Eq.(\ref{4-modeQD2}), we notice that one can essentially treat ${\bf D}$ as a constant vector in the limit $\w/\mu\to0$. Thus, in Eq.(\ref{4-modeQD1})$, {\bf Q}$ acquires an extra precession around the ${\bf D}$ vector. This precession is essentially around the $z$ axis, and now allows the $y$ component of ${\bf Q}$ to evolve as well. The vectors ${\bf X}$ and ${\bf Y}$ also acquire similar precessions around ${\bf D}$, but each with a different precession frequency. As these frequencies are not all identical, there is no ``co-rotating'' frame where all the effects of these additional precessions can be completely removed. 

\begin{figure}[!t]
\includegraphics[width=0.44\textwidth,height=0.25\textwidth]{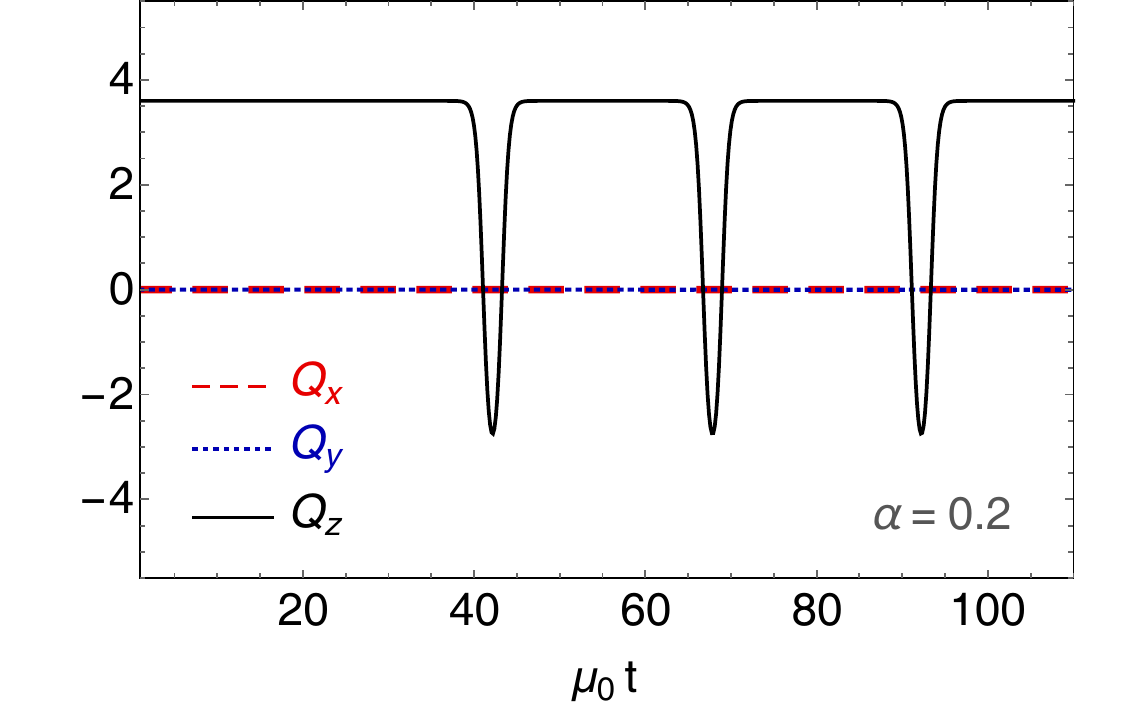}
\caption{Dynamics of the components of ${\bf Q}$ for a neutrino-antineutrino asymmetry $\alpha=0.2$. The parameters are chosen to be $\w/\mu_0=10^{-5}$, $\vartheta_0=10^{-2}$ and $c=0.5$.}
\label{fig8}
\end{figure}

Taking a derivative of Eq.(\ref{4-modeQD1}), one gets
\begin{align}
\ddot{\bf Q}        =-\mu^2\,c\,(1-c)\biggl[|{\bf Q}_0|^2-{\bf Q}\cdot{\bf Q}\biggr]{\bf Q}+\frac{\mu}{2}(3-c)\,{\bf D}\times\dot{\bf Q}\,\nonumber\\
+ \frac{\mu^2}{2}(1+c)\left[({\bf D}\times{\bf X})\times{\bf Y}+\frac{3+c}{2}\,{\bf X}\times({\bf D}\times{\bf Y})\right]\,.\nonumber\\
\end{align}
The second term on the r.h.s of the first line represents the action of a approximately constant magnetic field ${\bf D}\approx(0,\,0,\,2\alpha)$ in the $z$ direction. The terms on the second line are approximately equal to $({\bf X}.{\bf Y}){\bf D}$, which act like a time-varying electric field in the $z$ direction. Despite these complications, the interpretation is not too difficult. For $\alpha=0$, the $Q_z$ already hovers close to its minimum around $-4$, but $|{\bf Q}|$ is constrained to be $\leq 4$. Now, with $\alpha\neq 0$, the only possible effect of these new terms can be that $Q_z$ becomes larger close to its minimum. This is exactly what is seen in Fig.\,\ref{fig8}; the dips become less deep and are sharper. Essentially, these electric and magnetic fields push the particle away from the minimum of the potential well.

\section{Summary and Outlook}
\label{sec:4}
In this paper, we studied the simplest toy model of a homogeneous system of neutrinos and antineutrinos that shows fast conversions. We find that, in the limit that the vacuum oscillation frequency $\w$ is much smaller than the neutrino potential $\mu$, the system is described by a particle moving in a quartic potential (and an external electric and magnetic field, if there is neutrino-antineutrino asymmetry). This simple classical mechanical problem can be solved exactly.
Most importantly, the potential offers a barrier as opposed to a well, if the angle of intersection of the beams is larger than $\pi/2$, which explains the dependence of fast conversions on the angular distribution of the beams. Onset of fast conversions corresponds to the particle rolling down the potential, thereby causing an instability. Using the action variable and its adiabatic invariance, we estimated the time-period of fast oscillation, both when $\mu$ is constant and when $\mu(t)$ varies with time. We gave numerical and semi-analytical evidence that the onset and waiting periods for the fast oscillations depend logarithmically on $\vartheta_0$ and ${\cal O}(\w/\mu)$. Finally, we argued how our results generalize to a situation when the number of neutrinos and antineutrinos is not same. In this case, the particle is also acted upon by an external electric and  magnetic field. 

We hope that these results provide some useful insight of the flavor dynamics associated with fast oscillations, that has so far only been understood in the linear regime or explored numerically. Hopefully, these insights will be useful to understand the physics of fast oscillations in more realistic models of neutrino flavor conversions in core collapse supernovae. 

\section*{Acknowledgements}
We thank Amol Dighe for helpful discussions. This work was partially funded through a Ramanujan Fellowship of the Dept. of Science and Technology, Government of India, and the Max-Planck-Partnergroup ``Astroparticle Physics'' of the Max-Planck-Gesellschaft awarded to B.D. The work of M.S. was supported by the Dept. of Atomic Energy, Government of India. This project has received partial support from the European Union's Horizon 2020 research and innovation programme under the Marie-Sklodowska-Curie grant agreement Nos.\,674896 and 690575.

\onecolumngrid
\appendix

\section{Full EoMs for Fast Conversions}
\label{sec:App}

\begin{figure*}
\includegraphics[width=0.40\textwidth,height=0.25\textwidth]{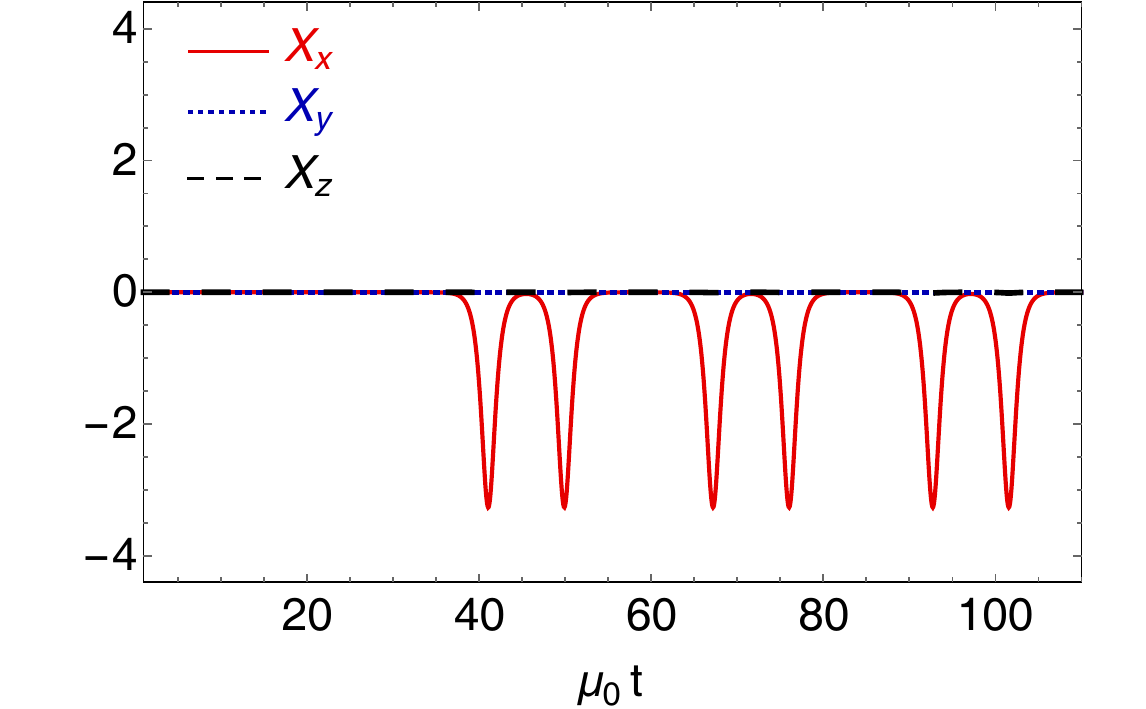}\,\, \includegraphics[width=0.40\textwidth,height=0.25\textwidth]{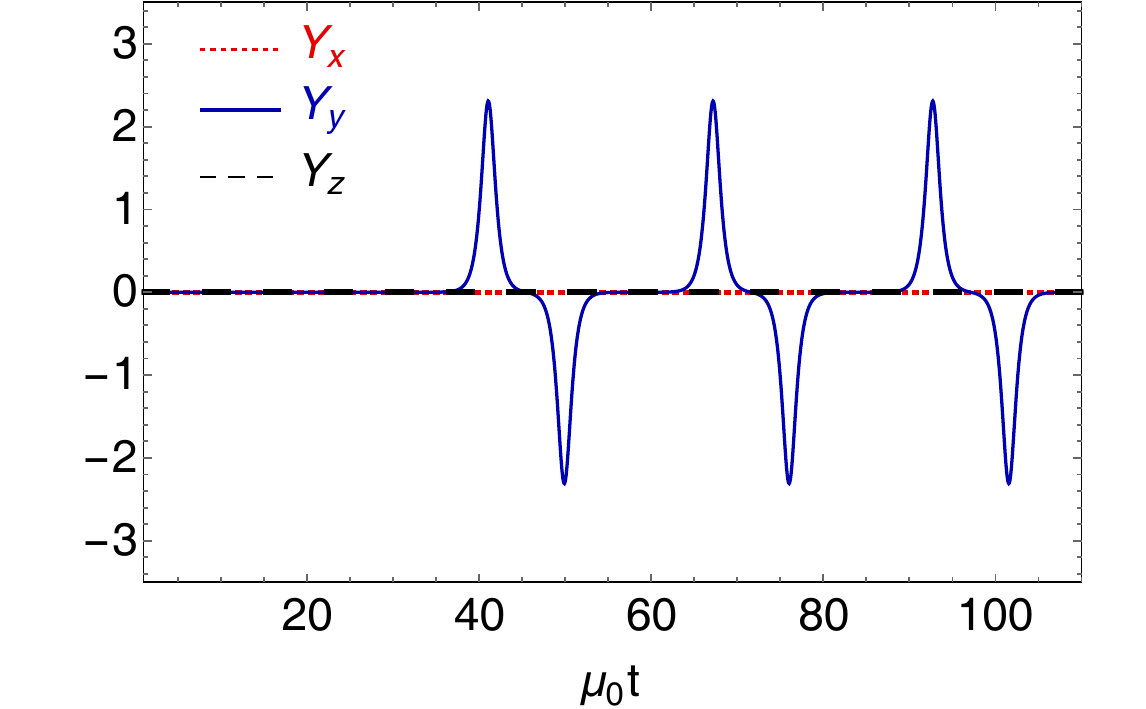}
\caption{Left: Evolution of ${\bf X}$. Right: Evolution of ${\bf Y}$. The parameters used here are $\w/\mu_0=10^{-5}$, $\vartheta_0=10^{-2}$ and $c=0.5$. }
\label{figA1}
\end{figure*}

In this appendix, for completeness, we provide the EoMs for the polarization vectors as well as the EoMs for ${\bf Q},\,{\bf X}$ and ${\bf Y}$ without dropping the subleading terms. The EoMs for the four polarization vectors are given by:
\begin{eqnarray} \label{4-modeP}
 \dot{\bf P}_{\bf L}               &=& \w {\bf B}\times{\bf P_L}+\mu\bigl[(1+c)\,{\bf P_R}-(1-c)\,\overline{{\bf P}}_{\bf L}-2\,\overline{{\bf P}}_{\bf R}\bigr]\times {\bf P_L}\,, \nonumber\\
 \dot{\bf P}_{\bf R}               &=& \w {\bf B}\times{\bf P_R}+\mu\bigl[(1+c)\,{\bf P_L}-(1-c)\,\overline{{\bf P}}_{\bf R}-2\,\overline{{\bf P}}_{\bf L}\bigr]\times {\bf P_R}\, , \nonumber\\
\dot{\overline{{\bf P}}}_{\bf L}   &=&-\w{\bf B}\times\overline{{\bf P}}_{\bf L}+\mu\bigl[(1-c)\,{\bf P_L}-(1+c)\,\overline{{\bf P}}_{\bf R}+2\,{\bf P}_{\bf R}\bigr]\times \overline{{\bf P}}_{\bf L}\, , \nonumber\\
\dot{\overline{{\bf P}}}_{\bf R}   &=&-\w {\bf B}\times\overline{{\bf P}}_{\bf R}+\mu\bigl[(1-c)\,{\bf P_R}-(1+c)\,\overline{{\bf P}}_{\bf L}+2\,{\bf P}_{\bf L}\bigr]\times \overline{{\bf P}}_{\bf R} \, ,
\end{eqnarray}
where $c\equiv\cos\theta$ is the cosine of the angle between the beams, as shown in Fig.\,\ref{fig1}. Using the definitions for ${\bf Q},\,{\bf D},\,{\bf X}$ and ${\bf Y}$ in Eqs.(\ref{redef1}\,-\,\ref{redef4}), we have already shown the evolution of ${\bf Q}$ in the main text and here, in Fig.\,\ref{figA1}, we show the evolution of ${\bf X}$ and ${\bf Y}$, for $\alpha=0$. We observe that while ${\bf X}$ develops only an $x$ component dominantly (and has a subleading $z$ component), the quantity ${\bf Y}$ only has a non-zero $y$ component. This can also be inferred by inspecting the EoMs. ${\bf D}$ remains very small and along the $y$ direction and we do not show it here.

In the process of identifying the above equations with the mechanical analog, the crucial approximation was to drop the subleading terms of frequency ${\cal O}(\w\mu)$ and smaller. These terms are manifest in the approximate second-order EoMs for ${\bf Q},\,{\bf X},$ and ${\bf Y}$ which, in the limit $\alpha=0$, can be arrived at by taking another time-derivative of Eqs.(\ref{4-modeD01}\,-\,\ref{4-modeD03}),
\begin{eqnarray}\label{4-modeDoubleder}
 \ddot{\bf Q}     &=& \frac{\mu}{2}(1+c)\,\Biggl[\mu\,c\biggl\{({\bf Y}\cdot{\bf Q}){\bf Y}-({\bf Y}\cdot{\bf Y}){\bf Q}\biggr\}-\frac{\mu}{2}(1-c)\biggl\{({\bf X}\cdot{\bf X}){\bf Q}-({\bf X}\cdot{\bf Q}){\bf X}\biggr\} \nonumber\\
                      & &\quad\quad\quad\quad+\w\left(\frac{3+c}{3-c}\right)\biggl\{({\bf Y}\cdot{\bf B}){\bf Y}-({\bf Y}\cdot{\bf Y}){\bf B}\biggr\}+
\w\left(\frac{2}{3-c}\right)\biggl\{({\bf X}\cdot{\bf X}){\bf B}-({\bf X}\cdot{\bf B}){\bf X}\biggr\}\Biggr]\,,\\  
\ddot{\bf X}  &=&\mu\,c\,\Biggl[\frac{\mu}{2}(1+c)\biggl\{({\bf Y}\cdot{\bf X}){\bf Y}-({\bf Y}\cdot{\bf Y}){\bf X}\biggr\}-\frac{\mu}{2}(1-c)\biggl\{({\bf Q}\cdot{\bf X}){\bf Q}-({\bf Q}\cdot{\bf Q}){\bf X}\biggr\}\, \nonumber\\
		      & &\quad\,\,+\w\left(\frac{2}{3-c}\right)\biggl\{({\bf Q}\cdot{\bf X}){\bf B}-({\bf Q}\cdot{\bf B}){\bf X}\biggr\}\Biggr]+
\w\left(\frac{3+c}{3-c}\right){\bf B}\times\dot{\bf Y}\,,\\   
\ddot{\bf Y}  &=&-\frac{\mu}{2}\,(1-c)\,\Biggl[\frac{\mu}{2}(1+c)\biggl\{({\bf X}\cdot{\bf X}){\bf Y}-({\bf X}\cdot{\bf Y}){\bf X}\biggr\}+\mu\,c\biggl\{({\bf Q}\cdot{\bf Y}){\bf Q}-({\bf Q}\cdot{\bf Q}){\bf Y}\biggr\} \nonumber\\
		      & &\quad\quad\quad\quad\quad+\w\left(\frac{3+c}{3-c}\right)\biggl\{({\bf Q}\cdot{\bf Y}){\bf B}-({\bf Q}\cdot{\bf B}){\bf Y}\biggr\}\Biggr]+
\w\left(\frac{2}{3-c}\right){\bf B}\times\dot{\bf X}\,.   
\end{eqnarray}
We remind that these equations are based on the assumption that ${\bf D}$ is approximately constant and negligible. Also, the apparently ${\cal O}(\mu^2)$ terms on the first line of the above equations contain subleading ${\cal O}(\w\mu)$ terms themselves.

Analogous to the closed set of equations and the Lagrangian governing ${\bf Q}$ given by Eq.(\ref{ModeQ1}), one can find the closed equation for ${\bf X}$ and ${\bf Y}$, each, by neglecting terms of order ${\cal O}(\w^2)$ and ${\cal O}(\w\mu)$ relative to ${\cal O}(\mu^2)$,
\begin{eqnarray}
 \ddot{\bf X}&=& \mu^2\,\frac{c(1-c)}{2}\biggl[|{\bf Q}_0|^2-\frac{(1+c)}{c}({\bf X}\cdot{\bf X})\biggr]{\bf X}\,,\\
 \ddot{\bf Y} &=& \mu^2\,\frac{c(1-c)}{2}\biggl[|{\bf Q}_0|^2-2\,\frac{(1+c)}{(1-c)}({\bf Y}\cdot{\bf Y})\biggr]{\bf Y}\,.
\end{eqnarray}
\begin{figure*}
\includegraphics[scale=0.35]{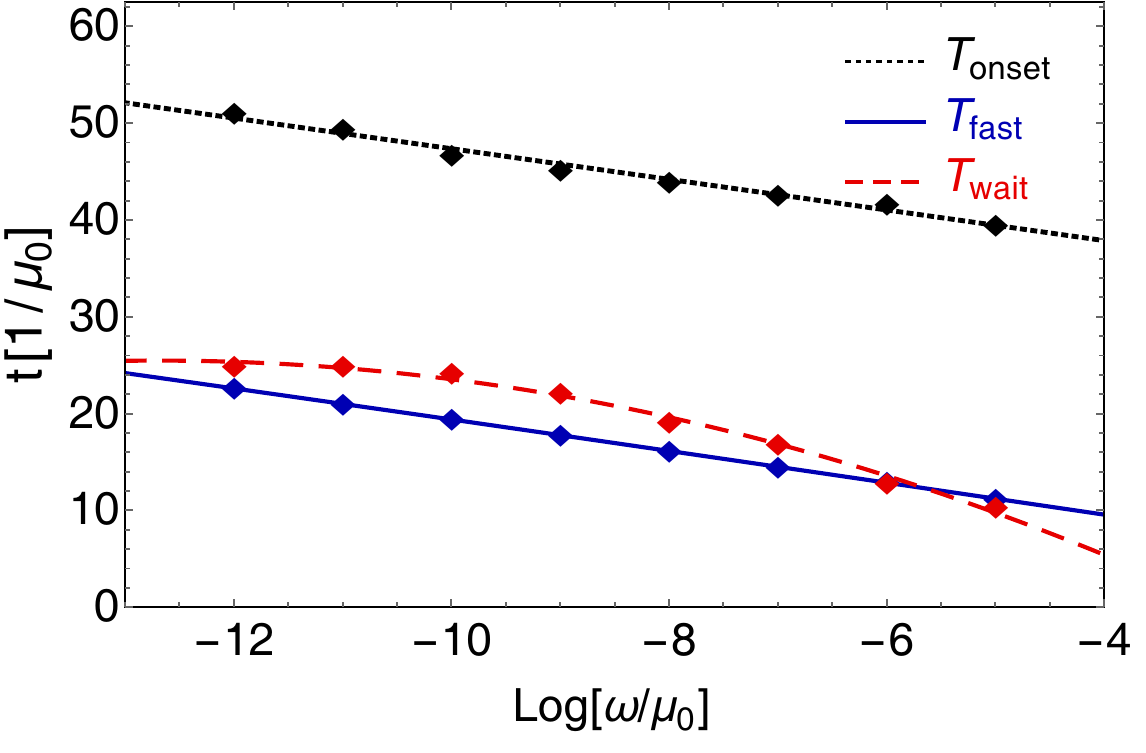}~~~~\includegraphics[scale=0.35]{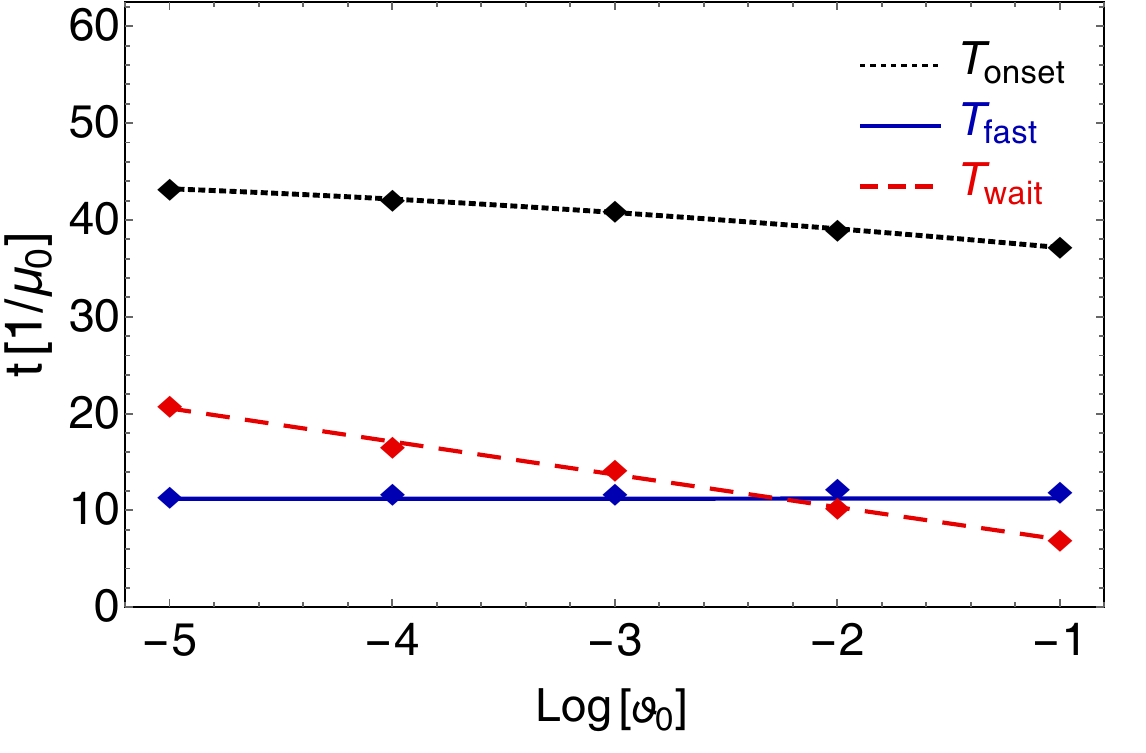}
\caption{Variation of the time periods $T_{\rm onset},\,T_{\rm fast}$ and $T_{\rm wait}$. Dots show the the data generated from simulations whereas the lines are the best fit curves through them. Here $\mu_0=10^{5}\,{\rm km}^{-1}$. Left: Variation with $\w/\mu_0$, for $\mu/\mu_0=1$ and $\vartheta_0=10^{-2}$. Right: Variation with $\vartheta_0$, for  $\w/\mu_0=10^{-5}$ and $\mu/\mu_0=1$.}
\label{figA2}
\end{figure*} 
In this $\alpha=0$ limit, the neglect of the subleading contributions of ${\cal O}(\w\mu)$ and smaller endows a spurious ${\bf Q}\to -{\bf Q}$ symmetry to Eq.(\ref{ModeQ1}). As a result, solving Eq.(\ref{ModeQ1}) leads to an evolution of ${\bf Q}$ that is exactly symmetric in $Q_z \leftrightarrow -Q_z$ (the onset and waiting times are equal to the fast oscillation time). Numerically however, we find that ${\bf Q}$ hovers longer around its initial position at the top, than it does at the bottom of the potential $V({\bf Q})$, as seen in Fig.\,\ref{fig2}. We believe that this slow-down is due to the neglect of subleading friction-like terms that arise at the same order as the terms necessary to seed the fast oscillation. Similar to how the onset period for the bipolar flavor pendulum depends on $\vartheta_0$, the time-scales for the fast oscillation, i.e., $T_{\rm onset}$, $T_{\rm fast}$, as well as $T_{\rm wait}$, depend logarithmically on these subleading parameters that seed the oscillations. In Fig.\,\ref{figA2}, we show the variation of $T_{\rm onset},\,T_{\rm fast}$ and $T_{\rm wait}$ with $\w/\mu_0$ and $\vartheta_0$, respectively, where $\mu_0=10^{5}\,{\rm km}^{-1}$ is the value of $\mu$ at the neutrinosphere. Clearly the time periods vary as $\mu^{-1}$ as shown in Fig.\,\ref{fig5}, but with logarithmic corrections proportional to $\left(\w/\mu_0\right)$ and $\vartheta_0$.

\twocolumngrid


\bibliographystyle{apsrev4-1}
\bibliography{fastosc.bib}

\end{document}